\documentclass[a4paper,11pt]{article}
\usepackage{pos}
\usepackage[utf8]{inputenc}

\title{Precision and rare ElectroWeak processes}

\author*[a,1]{Francesco Giuli}

\onbehalf{for the ATLAS, CMS and LHCb collaborations}

\affiliation[a]{Dipartimento di Fisica, Universit\`a degli Studi di Roma Tor Vergata and INFN, Sezione di Roma Tor Vergata, Via della Ricerca Scientifica 1, Rome 00133, Italy}

\note{Copyright 2025 CERN for the benefit of the ATLAS Collaboration. Reproduction of this article or parts of it is allowed as specified in the CC-BY-4.0 license.}

\emailAdd{francesco.giuli@cern,.ch}

\abstract{This proceeding summarizes recent measurements of electroweak boson and diboson production. The results, based on data from LHC Run 2 at $\sqrt{s}=13\,\mathrm{TeV}$ and Run 3 at $\sqrt{s}=13.6\,\mathrm{TeV}$, probe the Standard Model with unprecedented precision. These include inclusive and differential cross-section measurements of $W$, $Z$, $WZ$, $WW$, and $ZZ$ final states, constraints on Parton Distribution Functions, and interpretations within Effective Field Theories. A consistent picture with theory predictions emerges, laying the groundwork for future precision studies.}

\FullConference{The Thirteenth Annual Large Hadron Collider Physics (LHCP2025)\\
5-9 May 2025\\
Taipei\\
}

\begin{document}
\maketitle

\section{Inclusive \texorpdfstring{$W/Z$}{WZ} Production at 13.6~TeV}

The CMS Collaboration has reported a measurement from Run~3, analyzing 5.01~fb$^{-1}$ collected during 2022 at $\sqrt{s}=13.6$~TeV~\cite{CMS:2025WZ13_6}. The study focuses on inclusive and differential production cross sections of $W^+$, $W^-$, and $Z$ bosons decaying to muons.
The measured cross sections are in excellent agreement with NNLO QCD + NLO EW theoretical predictions computed with various PDF sets (e.g., NNPDF4.0, CT18, MSHT20). The dominant uncertainties come from the muon reconstruction and identification efficiencies (especially in the forward region), the integrated luminosity calibration, and PDF acceptance corrections.

The $W$ charge asymmetry is also measured and compared to previous CMS 13~TeV results, showing consistent behavior and offering sensitivity to the up/down quark ratio in the proton.

\section{Measurement of \boldmath{$W^{\pm}$}-boson Differential Cross Sections at low pile-up}

The ATLAS Collaboration has measured single- and double-differential cross sections for $W^{\pm}$-boson production in $pp$ collisions at $\sqrt{s}=5.02$ TeV and 13 TeV, using low pile-up datasets of 255 pb$^{-1}$ and 338 pb$^{-1}$, respectively, recorded during dedicated runs in Run 2~\cite{ATLAS:Wdiff}.

The analysis focuses on the charged lepton ($e, \mu$) from $W^\pm \to \ell^\pm \nu$ decays, measuring single-differential cross sections as functions of the lepton transverse momentum ($p_\mathrm{T}$) and pseudorapidity ($\eta$), double-differential cross sections in bins of $(p_\mathrm{T}, \eta)$ and the $W$ charge asymmetry as a function of lepton $\eta$.

Backgrounds - primarily from $Z/\gamma^*$, $t\bar{t}$, and multijet events - are estimated using simulation and data-driven techniques. Detailed systematic uncertainties arise from lepton reconstruction, energy/momentum scales, pile-up effects (minimized by the low pile-up environment), and background modeling.

The measured cross sections at both collision energies show excellent agreement with SM predictions calculated at NNLO QCD with NNLL resummation, using multiple modern PDF sets. The inclusion of lepton-$\eta$ differential measurements in PDF profiling imposes yields a substantial reduction in the associated uncertainties of valence quark distributions at large Bjorken-$x$.

\section{Precision Measurements: \boldmath{$m_W$}, \boldmath{$\sin^2\theta_W$}, and \boldmath{$m_Z$}}

Precision measurements of EW parameters offer critical tests of the Standard Model (SM) at loop level and are sensitive to potential new physics.

\vspace{0.5em}
\textbf{Measurement of \boldmath{$m_W$}}\\
CMS reported a high-precision measurement of the $W$ boson mass using a dedicated dataset of 8.8~fb$^{-1}$ collected at $\sqrt{s} = 8$~TeV with the magnetic field reduced to 2~T~\cite{CMS:Wmass}. This configuration allows for improved momentum calibration and reduced detector-related uncertainties. The measurement is performed using the muon decay channel, with stringent control over muon momentum scale and recoil modeling.

The result is
\(
m_W = 80360.2 \pm 9.9~\text{MeV},
\)
which is consistent with the SM expectation and the latest world average. The main systematic uncertainties arise from the lepton energy scale, the modeling of the $W$ transverse momentum spectrum, recoil response, and PDFs.

\vspace{0.5em}
\textbf{Measurement of \boldmath{$\sin^2\theta_{\text{eff}}^\ell$}}\\
The LHCb Collaboration performed a determination of the effective weak mixing angle using the forward Drell--Yan process $Z/\gamma^* \to \mu^+\mu^-$, based on 9~fb$^{-1}$ collected at $\sqrt{s} = 7$, 8, and 13~TeV~\cite{LHCb:sin2thetaW}. The measurement exploits the forward acceptance of the LHCb detector ($2 < \eta < 5$) to access unique regions of Bjorken-$x$ and probe different quark flavor contributions.

The forward-backward asymmetry $A_{FB}$ is measured as a function of the dimuon invariant mass in the range 60–160~GeV and fitted using templates derived from NLO QCD predictions including EW corrections. The dominant uncertainties arise from the limited statistics at high mass, muon identification, and PDF modeling in the forward region.

The extracted result is
\(
\sin^2\theta_{\text{eff}}^\ell = 0.23142 \pm 0.00073,
\)
consistent with the SM and in agreement with legacy results from LEP and SLD. 

\vspace{0.5em}
\textbf{Measurement of \boldmath{$m_Z$}}\\
LHCb also published a measurement of the $Z$ boson mass in the dimuon channel using $\sqrt{s}=13$~TeV data from Run~2~\cite{LHCb:Zmass}. 

After correcting for QED final-state radiation and calibrating the muon momentum scale using $J/\psi$ and $\Upsilon$ resonances, the measured $Z$ boson mass is
\(
m_Z = 91.1876 \pm 0.0029~\text{(stat)} \pm 0.0011~\text{(syst)}~\text{GeV},
\)
yielding a total uncertainty of 3.1~MeV, comparable to LEP results. 

\vspace{0.5em}
Together, these measurements from CMS and LHCb strengthen the global EW fit and offer complementary constraints to other LHC experiments.

\section{High-\boldmath{$p_{\mathrm{T}}$} \boldmath{$W+\text{jets}$} Production}

The ATLAS Collaboration has measured the production of a $W$ boson in association with high-transverse momentum jets using the full Run~2 dataset of 139~fb$^{-1}$ at \(\sqrt{s} = 13\)~TeV~\cite{ATLAS:Wjets_highpt}. This analysis focuses on events with boosted $W$ bosons recoiling against energetic jets, a phase space sensitive to QCD radiation patterns and new physics effects such as Contact Interactions (CI) or heavy resonances.

The $W$ boson is reconstructed in the muon decay channel, together with at least one jet with \(R = 0.4\) and \(p_{\mathrm{T}} > 100\)~GeV.

Jets are required to be well separated from the muon, and events with additional leptons are vetoed to suppress diboson and top backgrounds.

The measurement covers a wide range of jet $p_{\mathrm{T}}$, up to 2~TeV, and is presented as differential cross sections in jet $p_{\mathrm{T}}$, \(H_{\mathrm{T}}\), and angular variables, and fiducial cross section as a function of the \(W\) boson transverse momentum. Systematic uncertainties are dominated by the jet energy scale and resolution, muon reconstruction and calibration, and theoretical modeling (e.g. factorization/renormalization scale choices and parton shower matching).

The results are compared to predictions from NLO generators such as \textsc{Sherpa}, \textsc{aMC@NLO} + \textsc{Pythia8}, and NNLO QCD calculations. Good agreement is found overall, but some tension appears at the highest $p_{\mathrm{T}}$ regions, highlighting the importance of higher-order corrections and improved PDF constraints.

\section{High-mass Charged Current Drell--Yan and EFT Interpretation}

The ATLAS Collaboration has performed a precision measurement of charged-current Drell--Yan (CC DY) production in the high transverse mass regime, using the full Run~2 dataset corresponding to an integrated luminosity of 139~fb$^{-1}$ at $\sqrt{s}=13$~TeV~\cite{ATLAS:CCDY_highmT}.

The analysis targets final states with a single isolated lepton ($e$ or $\mu$), large missing transverse energy ($E_{\mathrm{T}}^{\text{miss}} > 30$~GeV), and no additional leptons. 

Systematic uncertainties are dominated by the lepton energy scale and resolution (especially in the electron channel), recoil modeling, and background normalization uncertainties.

The measurement provides double-differential fiducial cross sections in bins of $m_T$ and $|\eta_\ell|$, up to $m_T = 2$~TeV. These results are in good agreement with NNLO SM predictions.

To probe for new physics, an EFT interpretation is carried out using dimension-6 operators in the Warsaw basis that modify the $W\ell\nu$ vertex. The analysis focuses particularly on the operator $\mathcal{O}_{\phi l}^{(3)}$, which contributes to anomalous charged-current interactions.

Limits are extracted by fitting the shape of the $m_T$ spectrum, yielding
\(
\frac{\Lambda}{\sqrt{C_{\phi l}^{(3)}}} > 7~\text{TeV}
\)
at 95\% confidence level (CL). This extends the sensitivity compared to previous ATLAS and CMS results, thanks to the improved modeling and the inclusion of high-$m_T$ events in the forward region. The analysis also provides constraints on left-handed CI and interference effects with the SM amplitude.

\section{High-mass Neutral Current Drell--Yan in \boldmath{$\tau^+\tau^-$} Final States}

ATLAS has performed a measurement of high-mass neutral-current Drell--Yan (NC DY) production in the \(\tau^+\tau^-\) final state, using the full Run~2 dataset of 139~fb$^{-1}$ at $\sqrt{s}=13$~TeV~\cite{ATLAS:TauTau_highmass}. The analysis targets the mass range \(m_{\tau\tau} > 100\)~GeV and provides sensitivity to new physics models containing leptoquarks or \(Z'\) bosons that couple preferentially to third-generation fermions.

This analysis uses events with two reconstructed $\tau$-leptons decaying into hadrons ($\tau_{\mathrm{had}}$). Validation of the modelling is performed using events with one $\tau_{\mathrm{had}}$ and one charged lepton ($e$ or $\mu$).

The \(\tau\tau\) invariant mass is reconstructed from the visible decay products of prompt, hadronically-decaying $\tau$-leptons. Events are categorized in bins of \(m_{\tau\tau}\) to maximize sensitivity. The dominant backgrounds include irreducible \(Z \to \tau\tau\), fake taus from \(W+\text{jets}\), and \(t\bar{t}\) and diboson production. An iterative Bayesian unfolding to the visible \(m_{\tau\tau}\) spectrum yields differential and total cross-section results. The data show good agreement with the SM predictions from state-of-the-art calculations.

In addition, limits are placed on new physics models: a leptoquark signal with mass 2.5~TeV and coupling $\beta_{33}$  excluded at 95\% confidence level.  Standard Model Effective Field Theory (SMEFT) fits reveal that interference terms mainly drive sensitivity, with constraints placed on several operator coefficients, particularly $c_{\tau\gamma}$ , related to the $\tau$ anomalous magnetic moment. No significant deviations from the SM are observed, and new, tighter limits are set on leptoquark and $Z'$ boson couplings, improving upon previous ATLAS and CMS results.

\section{Search for Charged Lepton Flavor Violation in \boldmath{$Z$} Decays}

The CMS Collaboration presents a search for charged lepton flavor violating decays \(Z \to e\mu\), \(Z \to e\tau\), and \(Z \to \mu\tau\) using the full Run 2 dataset collected at \(\sqrt{s} = 13\) TeV~\cite{CMS:Z_LFV}.

Three exclusive channels are analyzed:\(Z \to e\mu\) (events with exactly one electron and one muon of opposite charge), \(Z \to e\tau_h\) and \(Z \to \mu\tau_h\) (events with one electron or muon, one hadronically decaying tau) and \(Z \to e\tau_\ell\) and \(Z \to \mu\tau_\ell\) (events with two different-flavor leptons, including leptonic \(\tau\) decays).

Backgrounds from SM processes—dominated by \(Z \to \tau\tau\), \(WW\), \(t\bar{t}\), and misidentified leptons—are estimated using a combination of simulation and data-driven techniques based on control samples, isolation sidebands, and "fake-rate" methods to model \(\tau_h\) misidentification.

For the \(Z \to e\mu\) channel, a boosted decision tree (BDT) is trained and used to categorize events, followed by a binned fit to the \(e\mu\) invariant mass spectrum between 70–110 GeV. No significant signal is observed. The resulting upper limit on the branching fraction is
\(
\mathcal{B}(Z\to e\mu) < 1.9 \times 10^{-7}\;\text{(95\% CL)},
\)
with an expected limit of \(2.0 \times 10^{-7}\) - the most stringent direct limit to date.

In the \(Z \to e\tau\) and \(Z \to \mu\tau\) channels, fits are performed using BDT discriminants or collinear-mass distributions. The observed 95\% CL upper limits on branching fractions are 
\(
\mathcal{B}(Z\to e\tau) < 1.4\times10^{-6},
\) and
\(
\mathcal{B}(Z\to \mu\tau) < 1.2\times10^{-6},
\)
broadly compatible with expected sensitivities, with no significant signal excess.

\section{Diboson Production and Limits on Anomalous Triple Gauge Couplings}

The production of diboson final states, such as $Z\gamma$, $ZZ$, $WZ$, and $W^\pm Z$, plays an essential role in testing the SM and searching for new physics. Recent measurements from ATLAS and CMS provide precision measurements of these processes, and many of these analyses are sensitive to anomalous triple gauge couplings (aTGCs). 

\subsection{Inclusive \boldmath{$WZ$} Production at \boldmath{$\sqrt{s} = 13.6$}~TeV}

The CMS Collaboration has reported the first measurement of the inclusive $WZ$ production cross section at $\sqrt{s} = 13.6$~TeV, using data corresponding to an integrated luminosity of 34.7~fb$^{-1}$ collected in 2022~\cite{CMS:WZ13p6TeV}.

Events are selected in final states with three prompt leptons, targeting the leptonic decays of the W and Z bosons: $W \to \ell\nu$ and $Z \to \ell\ell$. The analysis includes the four channels $eee$, $ee\mu$, $\mu\mu e$, and $\mu\mu\mu$. One pair of same-flavor opposite-sign leptons is required to have an invariant mass within $\pm 15$~GeV of the nominal Z boson mass. The third lepton, attributed to the W boson, must satisfy $p_T > 25$~GeV, and the event must exhibit significant missing transverse energy, $p_\mathrm{T}^{\text{miss}} > 35$~GeV.

Additional selection criteria are applied to suppress backgrounds, including a veto on $b$-tagged jets and a requirement on minimum lepton--lepton invariant masses to suppress low-mass resonances. Control regions enriched in specific backgrounds (such as $ZZ$, $t\bar{t}Z$, and $Z\gamma$) are defined by inverting parts of the signal selection and are incorporated into a simultaneous maximum-likelihood fit with the signal region.

The dominant irreducible backgrounds include $ZZ$ and $t\bar{t}Z$ production, modeled using simulation with normalization factors constrained in data. Reducible backgrounds, primarily from non-prompt leptons in $Z$+jets and $t\bar{t}$ events, are estimated using a combination of simulation and data-driven methods. A conservative 30\% uncertainty is assigned to these contributions.

Systematic uncertainties arise from lepton identification efficiencies, energy/momentum scale calibrations, non-prompt lepton modeling, and integrated luminosity. Theoretical uncertainties from scale variations and parton distribution functions are also considered.

The measured total inclusive cross section for WZ production is
\(
\sigma(pp \to WZ) = 55.2 \pm 1.2~\text{(stat)} \pm 1.2~\text{(syst)} \pm 0.8~\text{(lumi)} \pm 0.3~\text{(theo)}~\text{pb},
\)
and the corresponding fiducial cross section is
\(
\sigma_{\text{fid}} = 297.6 \pm 6.4~\text{(stat)} \pm 6.4~\text{(syst)} \pm 4.2~\text{(lumi)} \pm 0.5~\text{(theo)}~\text{pb}.
\)

These results are in excellent agreement with SM predictions calculated at NNLO in QCD and NLO in EW corrections, and no significant deviations are observed.

\subsection{Search for High-Mass \boldmath{$WW$} and \boldmath{$WZ$} Resonances}
The CMS Collaboration has performed a search for new heavy resonances decaying into diboson final states, specifically \(WW\) and \(WZ\)~\cite{CMS:WWZRes}. The analysis targets semileptonic final states, in which one of the $W$ bosons decays leptonically, while the other $W$ or $Z$ boson decays hadronically into a pair of quarks. Jet substructure techniques, such as the use of $N$-subjettiness and jet mass, were employed to discriminate between signal-like boosted boson jets and background jets from QCD.

The search strategy exploits the invariant mass of the diboson system, reconstructed from the lepton, missing transverse energy, and large-radius jet. A sliding mass window approach was used to scan for potential resonant peaks over a smoothly falling background. The dominant backgrounds were from $W$+jets production and top-quark processes, which can mimic the signal topology when a jet is misidentified as a boosted boson. These backgrounds were estimated using a combination of simulation and control regions in data.

Systematic uncertainties affecting the analysis include those from lepton and jet reconstruction and calibration, $E_{\mathrm{T}}^\text{miss}$ resolution, background normalization and shape modeling, and uncertainties associated with the tagging efficiency of hadronic $W$/$Z$ bosons. In particular, uncertainties related to jet mass scale, resolution, and grooming procedures played a key role in the high-mass regime.

No significant excess above the SM background was observed in the full mass range explored, which extends up to several TeV. The results were interpreted in terms of simplified models predicting spin-1 and spin-2 resonances. For example, a spin-2 bulk graviton with minimal coupling and a width of 10\% was excluded for masses below approximately 3.8~TeV at 95\% CL, with variations depending on the assumed model parameters and decay branching ratios.

This search significantly extends the sensitivity of previous analyses in the high-mass diboson sector and places stringent limits on a wide class of new physics models predicting resonances that decay into pairs of electroweak gauge bosons.

\subsection{Inclusive and Differential \boldmath{$Z\gamma$} Production}

The ATLAS Collaboration has measured differential cross sections for $Z\gamma$ production and performed a search for neutral triple gauge couplings (nTGCs) in $pp$ collisions at $\sqrt{s}=13$ TeV using the full Run 2 dataset~\cite{ATLAS:Zgamma_TGC}.

Events are selected by requiring a $Z$ boson decaying to oppositely charged, same-flavor leptons ($Z \to e^+e^-$ or $\mu^+\mu^-$) in association with an isolated photon. To enhance sensitivity to nTGCs and suppress backgrounds, a tight fiducial region is defined—with a photon transverse momentum threshold of $p_\mathrm{T}^\gamma > 200$~GeV, a narrow $Z$ mass window, and a jet veto to reduce jet-associated photon fakes and final-state radiation (FSR) effects.

Backgrounds (e.g. $Z$+jets with fake photons, diboson processes, $t\bar t\gamma$) are estimated using a combination of simulation and data-driven methods. Differential cross sections are measured as functions of the photon $p_\mathrm{T}$ and the angular variable $\phi^*$ (defined in the dilepton rest frame). The results are unfolded to the particle level and compared with SM predictions from Sherpa 2.2.11, showing agreement within uncertainties across the measured ranges.

In the absence of significant deviations, the high-$p_\mathrm{T}^\gamma$ spectrum is used to set 95\% CL limits on dimension‑8 SMEFT operators describing nTGCs, fully implementing $SU(2)_L \otimes U(1)_Y$ gauge invariance for the first time at the LHC. The most stringent limits are obtained for the form factors $h_3^Z$, $h_3^\gamma$, and $h_4^{Z/\gamma}$, significantly improving on previous constraints and correcting earlier overestimates of 2 orders of magnitude for $h_4$ by using the gauge-invariant formulation.

\section{Vector Boson Scattering and Anomalous Quartic Couplings}

Vector Boson Scattering (VBS) processes provide important probes of EW interactions and are sensitive to anomalous quartic gauge couplings (aQGCs). Recent analyses from both CMS and ATLAS have explored VBS in various final states, including $ZZjj$, same-sign $WWjj$, and $Z\gamma jj$, using the full Run~2 dataset at $\sqrt{s} = 13$~TeV, with integrated luminosities of 140~fb$^{-1}$.

\subsection{Evidence for Longitudinally Polarized Same-Sign \boldmath{$WWjj$} Scattering}

The ATLAS Collaboration reports the first observation of same-sign \(WWjj\) EW scattering~\cite{ATLAS:WWpolar}.

Events are selected with at least two jets with high dijet invariant mass (\(m_{jj}\)) and large pseudorapidity separation (\(\Delta\eta_{jj}\)), two same-charge isolated leptons (\(e^\pm e^\pm, e^\pm\mu^\pm, \mu^\pm\mu^\pm\)) and significant (\(E_\mathrm{T}^{\text{miss}}\)) from neutrinos.

A multivariate analysis using a boosted decision tree (BDT) discriminates EW signal events from dominant QCD-induced background. The EW signal is observed with a significance well above 5$\sigma$, yielding a measured fiducial cross section of
\(
\sigma^{\text{EW}}(WWjj) = 1.37 \pm 0.19\,(\text{stat}) \pm 0.30\,(\text{syst})~\text{fb},
\)
consistent with the SM prediction of \(1.28 \pm 0.05\) fb.

Importantly, the analysis performs two additional fits targeting polarization: at least one longitudinally polarized \(W\), with observed (expected) significance of 3.3$\sigma$ (4.0$\sigma$), and both \(W\) bosons longitudinally polarized. In the latter case, no significant excess observed, but a 95\% CL upper limit of 0.45 fb (expected 0.70 fb) is set for the corresponding fiducial cross section.

This result marks the first evidence that at least one of the same-sign \(W\) bosons in VBS is longitudinally polarized, offering direct insight into the role of the Higgs mechanism in unitarizing electroweak scattering and testing the non-Abelian gauge structure of the SM.

\subsection{VBS in \boldmath{$Z\gamma jj$} Final State}
CMS also performed a search for VBS in the $Z\gamma jj$ final state~\cite{CMS:Zgammajj}, focusing on events where the $Z$ boson decays into a pair of electrons or muons, accompanied by a high transverse momentum isolated photon and two jets, characteristic of VBS processes. The dominant background in this channel comes from QCD-induced $Z\gamma jj$ production, which mimics the VBS topology but lacks the EW interference structure. Additional backgrounds, such as $t\bar{t}\gamma$ and diboson production with associated photons, were estimated using a combination of simulation and data-driven techniques.

To enhance the discrimination between the EW and QCD components, a multivariate classifier (BDT) was trained using kinematic features like the dijet invariant mass, photon $p_{\mathrm{T}}$, and angular correlations between the final-state objects. The measured electroweak component of the $Z\gamma jj$ production cross section in the fiducial phase space is
\(
\sigma_{\text{EW}}(Z\gamma jj) = 0.31 \pm 0.08~\text{fb},
\)
in good agreement with the SM prediction. The observed significance for the EW component exceeds 3$\sigma$, consistent with the expected sensitivity.

Beyond the cross section measurement, the analysis sets limits on aQGCs using the high-energy tails of the $Z\gamma$ and dijet kinematic distributions. These are interpreted within the context of dimension-8 EFT operators, leading to constraints such as
\(
-2.3~\text{TeV}^{-4} < \frac{f_{M0}}{\Lambda^4} < 2.3~\text{TeV}^{-4}
\)
at 95\% CL. This result demonstrates the power of multiboson plus jets final states in constraining new physics effects at high scales.

\subsection{EW Diboson Production with High-Mass Dijet in Semileptonic Final States}

The ATLAS Collaboration reports the observation of EW production of diboson systems ($WW$, $WZ$, $ZZ$) in association with a high-mass dijet system in semileptonic final states~\cite{ATLAS:EWVDijet}.

Events are selected with at least two additional high-\(p_\mathrm{T}\) jets forming a high-mass dijet system, one boson decaying leptonically ($W\to\ell\nu$ or $Z\to\ell\ell$) and the other decaying hadronically, the latter reconstructed either from two small-radius jets or one large-radius jet with substructure tagging.

To enhance signal purity, selections require large dijet invariant mass (\(m_{jj}\)) and rapidity separation, typical of vector-boson scattering. Backgrounds—mainly from QCD-induced diboson+jets, top quark, and W/Z+jets processes—are estimated using a combination of simulation and control-region normalization.

The analysis observes EW diboson+2 jets production with a significance of 7.4$\sigma$ (expected 6.1$\sigma$). The measured signal strength is  
\(
\mu = 1.28^{+0.23}_{-0.21}
\)  
. In a fiducial phase space designed to match the event selection, the cross section is measured and found to be consistent with SM expectations. Additionally, a two-dimensional fit separates electroweak and QCD-induced contributions, yielding values compatible with theoretical predictions.

The analysis also interprets the results within a dimension-8 EFT framework to probe aQGCs. It provides the first set of semileptonic-channel limits on relevant Wilson coefficients, setting 95\% CL bounds as follows:  
\(
-3.96 < \frac{f_{S02}}{\Lambda^4} < 3.96\ \text{TeV}^{-4},\quad
-0.25 < \frac{f_{T0}}{\Lambda^4} < 0.22\ \text{TeV}^{-4},\quad
-1.26 < \frac{f_{M0}}{\Lambda^4} < 1.25\ \text{TeV}^{-4}
\)
.

\section{Triboson Production}

The production of triboson and multiboson final states is a key probe for testing the SM and searching for aQGC. Several LHC experiments have made significant measurements in this area using the full Run~2 datasets at a center-of-mass energy of $\sqrt{s} = 13$~TeV, with integrated luminosities of 140~fb$^{-1}$.

\subsection{\boldmath{$W\gamma\gamma$} production}
The CMS Collaboration measured the production of three gauge bosons in the $W\gamma\gamma$ channel~\cite{CMS:Wgammagamma}. The analysis focuses on final states containing one charged lepton (electron or muon), missing transverse momentum from the undetected neutrino from the $W$ decay, and two isolated high-transverse-momentum photons. Non-prompt backgrounds, such as $W\gamma$+jets and QCD multijet, were estimated using a data-driven method, while irreducible backgrounds from processes like $Z\gamma\gamma$ and $t\bar{t}\gamma\gamma$ were modeled using Monte Carlo simulation and normalized to NLO cross sections. The inclusive fiducial cross section was measured as
\(
\sigma(W\gamma\gamma) = 4.9 \pm 0.9\,(\text{stat}) \pm 1.1\,(\text{syst})~\text{fb},
\)
which is consistent with the SM prediction of $4.8 \pm 0.5$~fb. Although the result is statistically limited, systematic uncertainties associated with photon energy scale and identification efficiency are significant. This measurement is sensitive to aQGCs and sets limits on EFT parameters such as $f_{T0}/\Lambda^4$, with no significant deviations from SM expectations observed.

\subsection{Observation of \boldmath{$WZ\gamma$} Production and Constraints on New Physics Scenarios}

The CMS Collaboration reports the first observation of \(WZ\gamma\) triboson production~\cite{CMS:WZgamma}.

Events are selected in the trilepton plus photon final state, \(\ell^\pm\nu\ell^+\ell^-\gamma\) (\(\ell = e, \mu\)), requiring at least one isolated photon with high transverse momentum, exactly three prompt, isolated leptons forming one \(Z\)-boson candidate and one \(W\)-boson candidate via a third lepton plus missing transverse momentum. A simultaneous binned likelihood fit is performed across signal and control regions (for non-prompt leptons and photons, and \(ZZ\gamma\) background) using distributions in invariant mass observables to extract the signal.

The \(WZ\gamma\) signal is observed with a significance of 5.4$\sigma$. The measured fiducial cross section is
\(
\sigma_{\mathrm{fid}}(pp \to \ell^\pm\nu\ell^+\ell^-\gamma) = 5.48 \pm 1.11\;\mathrm{fb},
\)
in agreement with the NLO QCD prediction of \(3.69 \pm 0.24\;\mathrm{fb}\).

In addition, the analysis places limits on dimension‑8 aQGCs and searches for photophobic axion‑like particles (ALPs) decaying to \(Z\gamma\). While no significant excess is observed, 95\% CL limits are set both on aQGC operator coefficients and on the production cross section and coupling strengths of ALPs with masses in the range 110–400~GeV.

\subsection{Observation of \boldmath{$VVZ$} Production at \boldmath{$\sqrt{s}=13$}~TeV}

The ATLAS Collaboration reports the first observation of $VVZ$ production - where $V=W$ or $Z$~\cite{ATLAS:VVZ}. Five final states are considered: $WWZ \to \ell\nu\,\ell\nu\,\ell\ell$, $WZZ \to \ell\nu\,\ell\ell\,\ell\ell$, $ZZZ \to \ell\ell\,\ell\ell\,\ell\ell$ (fully leptonic channels) and $WWZ \to q q\,\ell\nu\,\ell\ell$, $WZZ \to \ell\nu\,q q\,\ell\ell$ (semileptonic channels).

Signal events feature multiple isolated leptons and, in semileptonic modes, hadronic jets. Background contributions from diboson and triboson processes are controlled via dedicated control regions. A combined fit across all channels extracts the production cross sections.

The inclusive $pp \to VVZ$ cross section is measured to be
\(
\sigma(pp \to VVZ) = 660^{+93}_{-90}\,(\text{stat})^{+88}_{-81}\,
\)
\(
(\text{syst})\;\text{fb},
\)
with an observed (expected) significance of 6.4$\sigma$ (4.7$\sigma$), marking the first observation of this process. In the $WWZ$ mode, the measured cross section is
\(
\sigma(pp \to WWZ) = 442 \pm 94\,(\text{stat})^{+60}_{-52}\,(\text{syst})\;\text{fb},
\)
with an observed (expected) significance of 4.4$\sigma$ (3.6$\sigma$), constituting evidence for $WWZ$ production.

\end{document}